\documentclass[twocolumn]{aastex63}
\usepackage{amsmath}

\usepackage{multirow}
\received{June 19th 2020}
\revised{August 31st 2020}
\accepted{Sept 30th 2020}
\submitjournal{AJ}

\shorttitle{ALMA Flux Calibration Accuracy}
\shortauthors{Francis et al.}
\graphicspath{{./}{figures/}}
\begin{document}

\title{On the accuracy of the ALMA flux calibration in the time domain and across spectral windows}

\correspondingauthor{Logan Francis}
\email{loganfrancis3@uvic.ca}

\author[0000-0001-8822-6327]{Logan Francis}
\affiliation{Department of Physics and Astronomy, University of Victoria,
3800 Finnerty Road, Elliot Building,
Victoria, BC, V8P 5C2, Canada}
\affiliation{NRC Herzberg Astronomy and Astrophysics,
5071 West Saanich Road,
Victoria, BC, V9E 2E7, Canada}
%\affiliation{loganfrancis3@uvic.ca}

\author[0000-0002-6773-459X]{Doug Johnstone}
\affiliation{NRC Herzberg Astronomy and Astrophysics,
5071 West Saanich Road,
Victoria, BC, V9E 2E7, Canada}
\affiliation{Department of Physics and Astronomy, University of Victoria,
3800 Finnerty Road, Elliot Building,
Victoria, BC, V8P 5C2, Canada}

\author[0000-0002-7154-6065]{Gregory Herczeg}
\affiliation{Kavli Institute for Astronomy and Astrophysics, Peking University, Beijing 100871, People's Republic of China}  
\affiliation{Department of Astronomy, School of Physics, Peking University, Beijing 100871, People's Republic of China}

\author[0000-0001-6492-0090]{Todd R. Hunter}
\affiliation{National Radio Astronomy Observatory, 520 Edgemont Rd, Charlottesville, VA 22903, USA}

\author[0000-0001-6307-4195]{Daniel Harsono}
\affiliation{Institute of Astronomy and Astrophysics, Academia Sinica, 
        No. 1, Sec. 4, Roosevelt Road, Taipei 10617, Taiwan, R.~O.~C.}




\begin{abstract}
A diverse array of science goals require accurate flux calibration of observations with the Atacama Large Millimeter/Submillimeter array (ALMA), however, this \added{goal} remains challenging due to the stochastic time-variability of the ``grid'' quasars ALMA uses for calibration.\deleted{The flux density of the grid quasars is regularly determined and cataloged by ALMA observations of solar system objects of known brightness; science observations are then calibrated by observing a grid quasar and using the catalog measurement nearest in time to extrapolate to the observing frequency.} In this work, we use 343.5 GHz (Band 7) ALMA Atacama Compact Array observations of four bright and stable young stellar objects over 7 epochs to independently assess the accuracy of the ALMA flux calibration and to refine the relative calibration across epochs. \replaced{The use of these four calibrators allows an unprecedented relative ALMA calibration accuracy of $\sim 3\%$}{The use of these four extra calibrators allow us to achieve an unprecedented relative ALMA calibration accuracy of $\sim 3\%$}. \replaced{Furthermore}{On the other hand}, when the observatory calibrator catalog is not up-to-date, the Band 7 data calibrated by the ALMA pipeline may have a flux calibration poorer than the nominal 10\%, which can be exacerbated by weather-related phase decorrelation when self-calibration of the science target is either not possible or not attempted. We also uncover a relative flux calibration uncertainty between spectral windows of \replaced{1}{0.8}\%, implying that measuring spectral indices within a single ALMA band \replaced{can be}{is likely} highly uncertain. We \added{thus} recommend various methods for science goals requiring high flux accuracy \replaced{a}{and} robust calibration, in particular, the observation of additional calibrators combined with a relative calibration strategy, and observation of solar system objects for high absolute accuracy. 
\end{abstract}


\section{Introduction}
\label{sec:intro}

%The accurate flux calibration of Atacama Large Millimeter/Submillimeter Array (ALMA) observations is important for a wide variety of science cases, but is challenging due to the paucity of bright and stable calibrator sources in the sub-mm/mm sky.

The accurate flux calibration of Atacama Large Millimeter/Submillimeter Array (ALMA) observations 
is crucial to a wide variety of science goals. For example, comparison of fluxes at different 
wavelengths, often using observations obtained at different times, leads to a spectral index that 
probes grain growth in disks  \citep[e.g.][]{ueda2020,pinilla2019}  and galaxies 
\citep[e.g.][]{sadaghiani2020,williams2019}.  In some cases, these spectral indices are measured 
within a band \citep{perez2020,lee2020}, which minimizes time-variability in the flux calibration 
but may introduce other uncertainties.  The flux calibration also affects results from programs 
that require accurate monitoring with time \citep[e.g][]{he2019,cleeves2017}, ratios of emission 
lines in different bands \citep[e.g.][]{flaherty2018, matra2017}, or comparison of fluxes from 
different objects in a survey \citep[e.g][]{tobin2020,ansdell2016}. For deep ALMA observations, 
multiple independently calibrated execution blocks are typically concatenated prior to imaging; 
accurate relative calibration improves the resulting image quality and self-calibration solutions 
\citep{andrews2018}.

Obtaining an accurate ALMA flux calibration, however, is exceptionally challenging due to the paucity of bright and stable calibrator sources in the sub-mm/mm sky. The most reliable calibrators are solar system objects, with large and predictable mm fluxes that are known to $\sim3-5\%$ \citep{butler2012}; unfortunately, solar system calibrators are located only in the plane of the ecliptic, and are therefore often not visible at the time of observation or are widely separated on the sky from the science target. As a result, ALMA observations are typically calibrated using ``grid'' calibrators - a collection of \replaced{47}{$\sim40$} mm-bright quasars distributed homogeneously across the sky \replaced{\citep{fomalont2014}}{\citep{almatech}}. Quasars, \added{however,} are variable in both flux and spectral index, \deleted{however,} so the grid calibrator fluxes must be determined by observation \replaced{with}{of} a solar system calibrator every 10-14 days in multiple ALMA bands \citep{almatech}. The grid calibrators are typically observed in both side bands of Band 3 (91.5 and 103.5 GHz) and Band 7 (343.5 GHz), with Band 6 (233 GHz) occasionally used in place of 7 when weather conditions are poorer. \replaced{To flux calibrate an ALMA observation,}{The procedure for ALMA flux calibration is thus the following:} a recently monitored grid source is \added{first} observed \citep{vankempen2014}, and its flux density in each spectral window is calculated by extrapolating from the nearest-in-time Band 3 measurement using a power law spectral index fit to the nearest-in-time pair of Band 3 and 7 measurements taken within 3 days of each other. This flux scale is then transferred to the phase calibrator --- typically a fainter quasar close to the science target --- and in turn to observations of the science target taken between phase calibrator scans. 

Despite the need for good flux calibration, there are few examples in the literature where the ALMA flux calibration accuracy is independently assessed. The time-variability of the ALMA grid calibrators has been quantified by the ALMACAL project for investigating quasar physics \citep{bonato2018}. The grid calibrators have been modelled using continuous time stochastic processes by \cite{guzman2019}, which can provide flux interpolation, forecasting, and uncertainty estimates taking into account the inherent time-variability. 

Two ALMA projects (PI: Logan Francis, project IDs 2018.1.00917.S, 2019.1.00475.S) are currently underway to precisely measure the sub-mm variability of 3 deeply embedded protostars in the Serpens Main molecular cloud \deleted{[distance: $436.0 \pm 9.2 $ pc \citep{ortizleon2017}]} at disk and inner envelope ($<  2000$ au) scales. 
In this work, we take advantage of the relative flux calibration strategy of these projects to independently test the accuracy of the flux scale determined during ALMA pipeline processing from the available grid calibrator data. The remainder of this paper is structured as follows: In Section \ref{sec:obs}, we describe the ALMA observations of our targets and data reduction, while in section \ref{sec:rel_cal} we present our relative calibration technique and analyze the pipeline calibration accuracy. In Section \ref{section:disc}, we discuss the impact of our findings on various science goals requiring good flux calibration accuracy and offer suggestions for best practices in reducing ALMA data. Section \ref{sec:conclusions} briefly summarizes the results of this work.

\section{Observations}
\label{sec:obs}

Our ALMA programs (2018.1.00917.S, 2019.1.00475.S) observe \replaced{4 possibly}{3 potentially} varying protostars (SMM 1, \replaced{SMM 2}{EC 53}, and SMM 10, \deleted{and SMM 2}) \added{\citep{johnstone2018,contreras2020,yoo2017}} and \replaced{an additional 4}{5 additional} young stellar object (YSO) calibrators (\added{SMM 2} SMM 9, SMM 4, \replaced{CAL 3}{SMM 3}, SMM 11) \added{in the Serpens Main molecular cloud \added{(distance: $436.0 \pm 9.2 $ pc, \citealt{ortizleon2017})}} at 343.5 GHz\deleted{to an RMS noise of 1 mJy}. \replaced{The stability or variability of our targets at outer envelope scales of $\sim 6000$ au is verified by contemporaneous sub-mm monitoring from the JCMT Transient Survey \citep{herczeg2017}}{These targets were selected based on their variability or stability as determined by the James Clerk Maxwell Telescope (JCMT) Transient Survey \citep{herczeg2017}. The ongoing Transient Survey monitors the brightness of YSOs in eight star forming regions at 450 $\mu$m and 850 $\mu$m (352.9 GHz) at a monthly or better cadence in order to identify changes in envelope brightness resulting from protostellar accretion variability. The JCMT resolution in Serpens Main is $\sim 6100$ au, however, and the bulk of the envelope response likely occurs at smaller scales \citep{Johnstone2013}. Our contemporaneous higher resolution ($\sim 1750$ au) ALMA observations thus provide a useful measurement of how protostellar envelopes respond to accretion variations. }

The brightness of all \added{our} YSO calibrators at the JCMT has remained stable over the first 4 years of the Transient Survey to $<3\%$. Since accretion outbursts of the YSO calibrators are possible, observing \replaced{4}{multiple calibrators} provides redundancy in the unlikely event that one becomes variable.

\begin{figure*}
    \centering
    \includegraphics[scale=0.6]{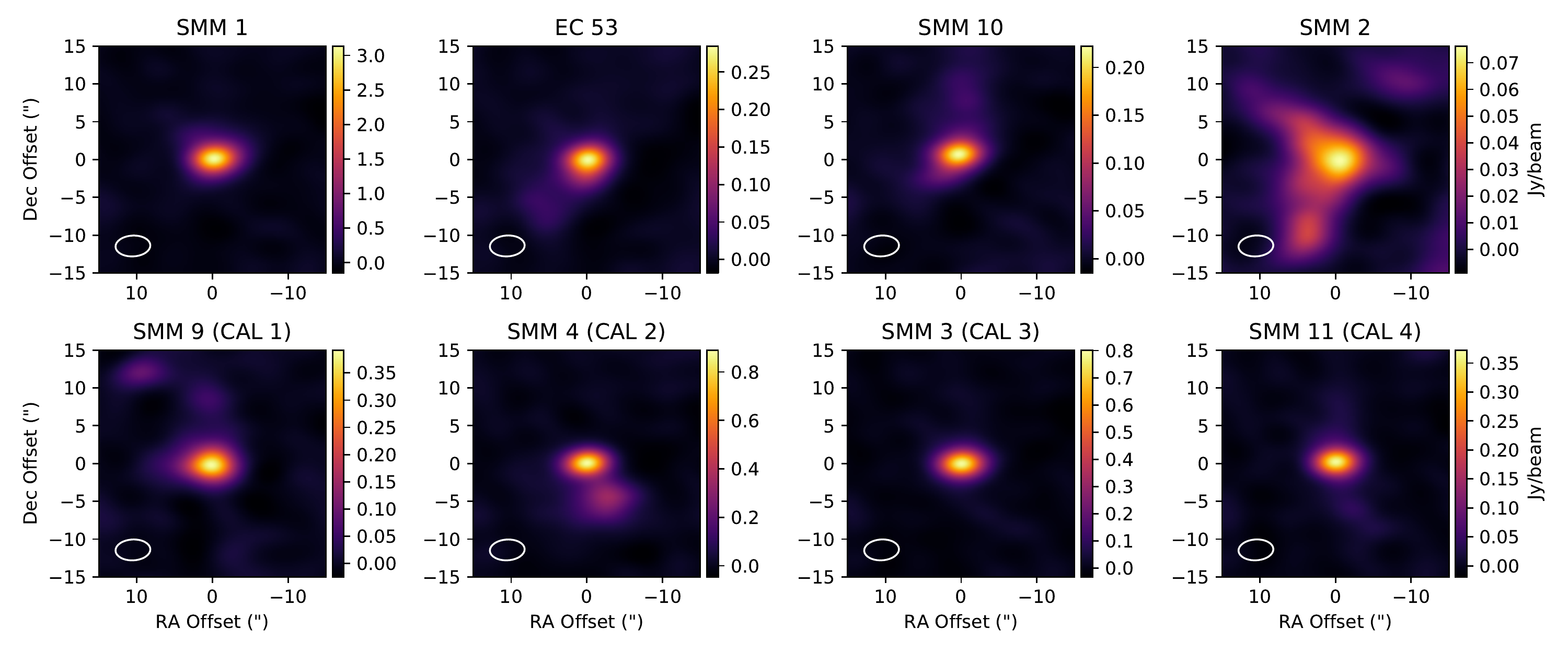}
    \caption{Deconvolved images of our targets constructed from concatenation of all epochs with the relative calibration in Section \ref{ssec:cont_cal} applied. \added{The CASA \texttt{tclean} task was used for image reconstruction with a robust weighting of 0.5.}}
        \label{fig:aca_deep_gallery}
\end{figure*}

All observations are taken with the stand-alone mode of the \added{Morita Array, otherwise known as the } Atacama Compact Array (ACA), a sub-array of ALMA consisting of twelve closely-spaced 7m diameter antennas. The ACA correlator is configured in \added{time division mode} with the default Band 7 continuum settings to provide 4 low-resolution spectral windows with 1.875GHz of bandwidth across 128 channels, for a total bandwidth of 7.5GHz. We have \replaced{to date} obtained 7 epochs of \added{observations of} our targets \added{as of July 2020} with a typical resolution of 4\arcsec ($\sim 1750$ au) \added{and RMS noise of $\sim1$ mJy}. \added{Names and coordinates of our targets are provided in Table \ref{tab:aca_coords}, while} the dates of observation and the flux and amplitude calibrators selected by the ALMA online system for our 7 epochs are listed in Table \ref{tab:aca_obs}. \replaced{Clean}{Deconvolved} images of our targets and calibrators constructed from concatenation of all available continuum data with the relative calibration discussed in section \ref{sec:rel_cal} applied are shown in Figure \ref{fig:aca_deep_gallery}. \added{While SMM2 is bright and stable in the JCMT Transient Survey, at the ACA resolution it is too faint and extended to obtain a useful calibration. We thus use SMM9, SMM4, SMM3, and SMM11 for relative calibration and hereafter refer to them as CAL 1-4.} Our target fluxes are $>300$ mJy (except SMM2), and as a result we can achieve a formal S/N $>300$, however, our images are dynamic range limited to a S/N of $\sim 100$. \added{We apply phase-only self-calibration to our observations, the procedure for and effects of which are discussed in \ref{ssec:self-cal}.} To avoid errors introduced by the deconvolution process in comparing target fluxes between observations, we perform our analysis in the $uv$-plane, where careful error analysis is more tractable. 

\begin{deluxetable*}{cccc}
\label{tab:aca_coords}
\tablecaption{ACA Science Targets and YSO Calibrators}
\tablehead{\colhead{Name} & \colhead{ALMA Name} & \colhead{Other mm Source Names} & \colhead{ACA field center (ICRS)}}
\startdata
Serpens SMM 1    &  Serpens\_Main\_850\_00 & Ser-emb 6, FIRS1    & 18:29:49.79 +01:15:20.4 \\
EC 53              &  Serpens\_Main\_850\_02 & Ser-emb 21          & 18:29:51.18 +01:16:40.4 \\
Serpens SMM 10 IR &  Serpens\_Main\_850\_03 & Ser-emb 12          & 18:29:52.00 +01:15:50.0 \\
Serpens SMM 2     &  Serpens\_Main\_850\_10 & Ser-emb 4 (N)       & 18:30:00.30 +01:12:59.4  \\ 
Serpens SMM 9 (CAL 1)    &  Serpens\_Main\_850\_01 & Ser-emb 8, SH2-68N  & 18:29:48.07 +01:16:43.7 \\
Serpens SMM 4 (CAL 2)    &  Serpens\_Main\_850\_08 &         -           & 18:29:56.72 +01:13:15.6 \\ 
Serpens SMM 3 (CAL 3)    &  Serpens\_Main\_850\_09 &         -           & 18:29:59.32 +01:14:00.5 \\
Serpens SMM 11 (CAL 4)   &  Serpens\_Main\_850\_11 &         -           & 18:30:00.38 +01:11:44.6
\enddata
%\tablecomments{}
\end{deluxetable*}

\begin{deluxetable}{cccc}
\label{tab:aca_obs}
\tablecaption{ACA Observing Dates and Pipeline Calibrators}
\tablehead{\colhead{Epoch} & \colhead{Date} & \colhead{Flux Calibrator} & \colhead{Phase Calibrator}}
\startdata
1 & 14-Oct-2018  & J1924-2914 & J1851+0035 \\
2 & 06-Mar-2019  & J1751+0939 & J1743-0350 \\
3 & 07-Apr-2019  & J1517-2422 & J1751+0939 \\
4 & 18-May-2019  & J1924-2914 & J1743-0350 \\
5 & 04-Aug-2019  & J1924-2914 & J1743-0350 \\
6 & 20-Sept-2019 & J1924-2914 & J1851+0035 \\
7 & 29-Oct-2019  & J1924-2914 & J1851+0035 \\
\enddata
%\tablecomments{}
\end{deluxetable}

\section{Relative Calibration of Atacama Compact Array data}
\label{sec:rel_cal}

Our 4\deleted{additional} YSO calibrators are monitored to allow the precise measurement of relative changes in the sub-mm flux of our science targets. Since JCMT monitoring has established that the brightness of the YSO calibrators is stable over 4 years to a level of $\sim$ 3-5\% \added{\citep{mairs2017}}, we can measure a relative calibration offset between epochs, providing a direct test of the ALMA calibration accuracy. Here we describe our relative calibration method, examine how the ALMA flux accuracy depends on the catalog, and quantify the accuracy of the ALMA flux calibration between spectral windows.

\subsection{Continuum Calibration}
\label{ssec:cont_cal}

We start our relative calibration process from the original pipeline-calibrated ALMA visibilities processed using CASA 5.6.1. For each target, we first construct dirty image cubes and extract the continuum by sigma-clipping of the spectra measured in a 3\arcsec diameter aperture centered on the brightest source; \added{typical $\sim90\%$ of the image cubes contain line-free continuum}. We then perform a $uv$-plane fit of a point source to the extracted continuum for each YSO calibrator and epoch assuming a zero (flat) spectral index across our spectral windows. While our sources \replaced{show a spectral index}{show spectral indices of $\sim$2-3} consistent with \added{envelope} thermal dust emission \added{\citep{ricci2010}}, we require only an average flux across the Band to perform a relative calibration between epochs, and moreover, we assume our YSO calibrators to be stable in spectral index. At the resolution of the ACA, most of our targets are approximately point sources with a fainter extended component (Figure \ref{fig:aca_deep_gallery}), so a point source model is generally sufficient for calibration. Future work will model the structure of each source in more detail.

For each of our 4 YSO calibrators, we measure the mean flux across all 7 epochs and calculate the ratio of the flux in each epoch to the mean light curve flux. \added{Through this, we obtain 4 independent estimates of the relative correction to the flux scale, whose distribution provides an estimate of the uncertainty in the relative calibration.} This technique follows that used for the JCMT Transient Survey \citep{mairs2017} and its application to interferometric data in \citet{francis2019}. The ``mean correction factor'' (MCF) for epoch $i$ and calibrator $j$ is thus:

\replaced{
\begin{equation}
    \mathrm{MCF}_{i,j}=\frac{\sum\limits_{i=1}^{7} F_{i,j}/7}{F_{i,j}}.
\end{equation}
}
{
\begin{equation}
    \mathrm{MCF}_{i,j}=\frac{\sum\limits_{i=1}^{7} F_{i,j}/7}{F_{i,j}};
\end{equation}
}
\added{while the uncertainty in the MCF is:

\begin{equation}
\label{eqn:mcf_err}
    \sigma_{MCF_{i,j}} = \frac{1}{N F_{i,j}} \left( \left(\sum_{\substack{l=1 \\ l\neq i}}^{7} F_{l,j} \right)^2\frac{\sigma_{F_{i,j}}^2}{F_{i,j}^2} 
        +   \sum_{\substack{l=1 \\ l\neq i}}^{7} \sigma_{F_{l,j}}^2 \right)^{1/2}.
\end{equation}

}
To determine a relative flux calibration factor (rFCF) for each epoch from our YSO calibrators, we take the average of the four MCFs for calibrators in that epoch, such that

\begin{equation}
    \mathrm{rFCF}_i= \sum\limits_{j=1}^{4}\mathrm{MCF}_{i,j}/4.
\end{equation}

\begin{figure}[htb]
    \centering
    \includegraphics[scale=0.93]{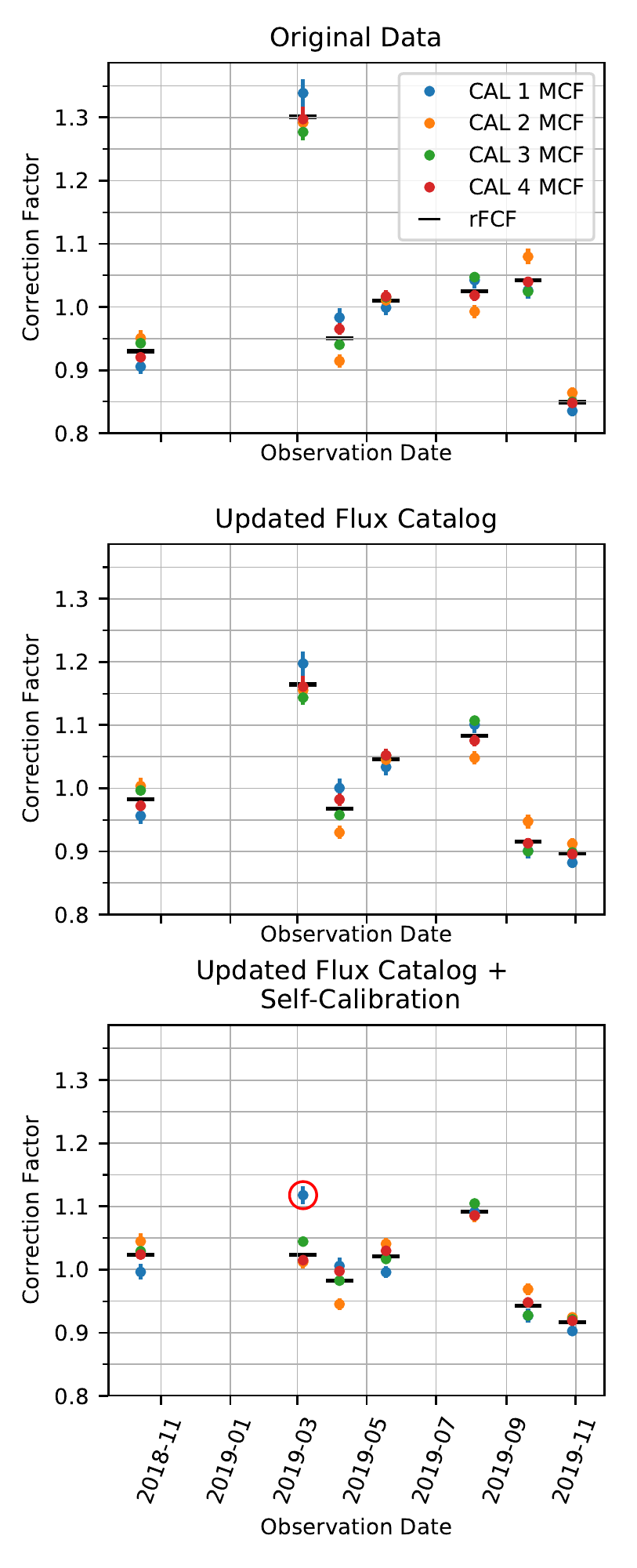}
    \caption{\replaced{MCFs and rFCFs}{Mean correction factors (MCFs) and relative flux calibration factors (rFCFs)} vs date for each of our YSO calibrators using visibilities from all spectral windows and assuming a flat spectral index. Top panel: MCFs calculated using the original grid calibrator fluxes. Middle panel: MCFs calculated using updated grid fluxes. Bottom panel: MCFs calculated with updated grid fluxes and phase self-calibration applied. The outlier MCF for CAL 1 in epoch 2 is circled in red.}
\label{fig:alma_MCFs}
\end{figure}

\replaced{In Figure \ref{fig:alma_MCFs}, we plot the MCF for each YSO calibrator and epoch.}{The MCFs and rFCFs thus calculated are listed in Table \ref{tab:mcfs_and_rfcfs}, and plotted vs the observing date in the top panel of Figure \ref{fig:alma_MCFs}}. For any epoch, the MCFs have a range $<7\%$ and standard deviation $<3\%$. The relative calibration accuracy is thus 3\% or better, which is unprecedented for ALMA data. We find the YSO rFCFs have a standard deviation of $14\%$ and range of $45\%$, in contrast with the expected nominal band 7 flux calibration accuracy of 10\% \citep{almaguide}. Notably, the second epoch requires a much larger rFCF, of $\sim$ 30\%.

\begin{deluxetable*}{cccccc}
\label{tab:mcfs_and_rfcfs}
\tablecaption{YSO Calibrator Correction Factors}
\tablehead{\multirow{2}{*}{Epoch} & \multicolumn{4}{c}{Mean Correction Factor (MCF)}  & \multirow{2}{*}{rFCF\tablenotemark{1}} \\ & \colhead{CAL 1} & \colhead{CAL 2} & \colhead{CAL 3} & \colhead{CAL 4}}
\startdata
1 & 1.00 $\pm$ 0.01 & 1.04 $\pm$ 0.01 & 1.029 $\pm$ 0.005 & 1.024 $\pm$ 0.007 & 1.02 \\
2 & 1.12 $\pm$ 0.01 & 1.01 $\pm$ 0.01 & 1.044 $\pm$ 0.006 & 1.015 $\pm$ 0.008 & 1.02 \\
3 & 1.01 $\pm$ 0.01 & 0.95 $\pm$ 0.01 & 0.982 $\pm$ 0.005 & 0.997 $\pm$ 0.008 & 0.98 \\
4 & 1.00 $\pm$ 0.01 & 1.04 $\pm$ 0.01 & 1.017 $\pm$ 0.003 & 1.030 $\pm$ 0.005 & 1.02 \\
5 & 1.09 $\pm$ 0.01 & 1.08 $\pm$ 0.01 & 1.105 $\pm$ 0.004 & 1.086 $\pm$ 0.006 & 1.09 \\
6 & 0.93 $\pm$ 0.01 & 0.97 $\pm$ 0.01 & 0.927 $\pm$ 0.005 & 0.948 $\pm$ 0.007 & 0.94 \\
7 & 0.90 $\pm$ 0.01 & 0.92 $\pm$ 0.01 & 0.921 $\pm$ 0.003 & 0.919 $\pm$ 0.005 & 0.92 
\enddata
\tablenotetext{1}{Relative Flux Calibration Factor}
\end{deluxetable*}

%With the outlier second epoch removed, the YSO rFCFs have  a standard deviation of $7\%$ and range of $19\%$, which is consistent with the expected ALMA flux calibration accuracy. 

\newpage
\subsection{Grid Calibrator Flux updates}

Because it represents a systematic offset in flux density, the large rRCF required in epoch 2 may result from a poor time interpolation if the grid calibrator catalog was out of date at the time of the original reduction. We thus queried the flux of each grid calibrator in April 2020 and compared its flux and spectral index with the values used by the pipeline in Table \ref{tab:pipeline}. Epochs 2 (06-Mar-2019) and 6 (20-Sept-2019) have changed by $\sim 20\%$, while smaller changes to epochs 3 and 4 of a few \replaced{\%}{percent} have also occurred. This large change in catalog flux is likely due to the inclusion of additional grid calibrator fluxes  in the ALMA catalog since the dates of the original reductions.

\begin{deluxetable*}{ccccccc}
\label{tab:pipeline}
\tablecaption{Original and updated pipeline flux calibrator values}
\tablehead{
\multirow{ 2}{*}{Epoch} & \multirow{ 2}{*}{Pipeline Run Date} & \multicolumn{2}{c}{Flux Density (Jy)} & \multicolumn{2}{c}{Spectral Index} & \multirow{ 2}{*}{Flux Density Change (\%)} \\
&  & \colhead{Original} & \colhead{Updated} & \colhead{Original} & \colhead{Updated} & 
}
\startdata
1 & 17-Oct-2018  & 2.65 $\pm$ 0.21 &  2.65 $\pm$ 0.09 & -0.609 $\pm$ 0.019  & -0.609 $\pm$ 0.019 & 0.0             \\
2 & 18-Apr-2019  & 2.08 $\pm$ 0.14 &  2.44 $\pm$ 0.07 & -0.590 $\pm$ 0.037  & -0.482 $\pm$ 0.014 & 17.3 $\pm$ 8.6  \\
3 & 18-Apr-2019  & 2.14 $\pm$ 0.13 &  2.22 $\pm$ 0.07 & -0.335 $\pm$ 0.025  & -0.306 $\pm$ 0.021 & 3.7  $\pm$ 7.1  \\
4 & 22-Sept-2019 & 2.13            &  2.17 $\pm$ 0.06 &       -             & -0.638 $\pm$ 0.044 & 1.8             \\
5 & 06-Sept-2019 & 2.15 $\pm$ 0.04 &  2.15 $\pm$ 0.04 & -0.668 $\pm$ 0.038  & -0.668 $\pm$ 0.038 & 0.0             \\
6 & 30-Sept-2019 & 2.40 $\pm$ 0.10 &  2.88 $\pm$ 0.06 & -0.642 $\pm$ 0.028  & -0.495 $\pm$ 0.039 & 20.0 $\pm$ 5.6  \\
7 & 19-Nov-2019  & 3.16 $\pm$ 0.06 &  3.16 $\pm$ 0.06 & -0.453 $\pm$ 0.008  & -0.453 $\pm$ 0.008 & 0.0               
\enddata
\tablecomments{All Fluxes are evaluated at center frequency of first spw of 336.495GHz. The original grid calibrator flux uncertainty and spectral index were not recorded by the pipeline in epoch 4.}
\end{deluxetable*}

With the updated pipeline values for the grid calibrator fluxes, we rescale the pipeline calibrated visibilities and compute the MCFs and rFCFs again \added{(middle panel of Figure \ref{fig:alma_MCFs})}. The magnitude of the rFCF in epoch 2 is now $\sim 15\%$, while the rFCFs overall now have a standard deviation of 9\% and range of 26\%, consistent with the nominal flux calibration accuracy. The reduced range of the rFCFs demonstrates the importance of using the most up-to-date catalog for achieving a good flux calibration. Further discussion and suggestions for best practices are given in Section \ref{section:disc}, while typical delays between observation of the flux calibrators and ingestion into the catalog are provided in section \ref{sec:alma_monitoring} of the Appendix.

\subsection{Effect of self-calibration}
\label{ssec:self-cal}
\added{
Weather conditions and instrumental effects can result in noisy or incorrect visibility phases, the extent of which can vary between observing epochs. Noisy phases may result in decorrelation and loss of flux during an observing scan \citep{brogan2018}, which would bias our mean correction factors to higher values. We thus apply  self-calibration to our YSO calibrators to assess its effect on our relative calibration. 

Three rounds of phase-only self-calibration were performed using solution intervals of a scan length, 20.2s, and 5.05s (an ACA integration is 1.01s in time division mode). Models of each source were constructed with the casa \texttt{tclean} task with a robust weighting of 0.5, and calibration solutions were allowed to vary between spectral windows. Repeating our uv-plane point souce fits, we find the calibrator fluxes to increase by a few percent for all but epoch 2, where the improvement was $\sim15\%$. The resulting MCFs and rFCFs are shown in the bottom panel of figure \ref{fig:alma_MCFs}; we exclude the MCF of CAL 1 in epoch 2 from the calculation of the rFCF as it as an outlier in its flux increase. We find the overall MCF standard deviation remains at $\sim3\%$, while the standard deviation and range of the rFCFs are now further reduced to 5\% and 17\% respectively; this is largely the result of the $\sim15\%$ in flux of the YSO calibrators in epoch 2, which pulls the rFCFs for the other epochs closer to 1. Phase self-calibration is thus important for relative calibration in order to avoid biasing of the flux rescaling. 
}
\newpage
\subsection{Calibration across spectral windows}
\label{ssec:norm_mcfs}

We repeat the process of obtaining our \replaced{MCFs and rFCFs}{mean correction factors} for each spectral window independently using the updated grid calibrator fluxes \added{and phase self-calibrated data}, thus allowing us to measure the accuracy of the ALMA flux calibration between spectral windows. As our YSO calibrator sources have a S/N $>300$ using data from all 4 spectral windows, the factor of 2 decrease in S/N resulting from use of a single window should not significantly affect our analysis. 

\begin{figure*}[htb]
    \centering
    \includegraphics[scale=0.95]{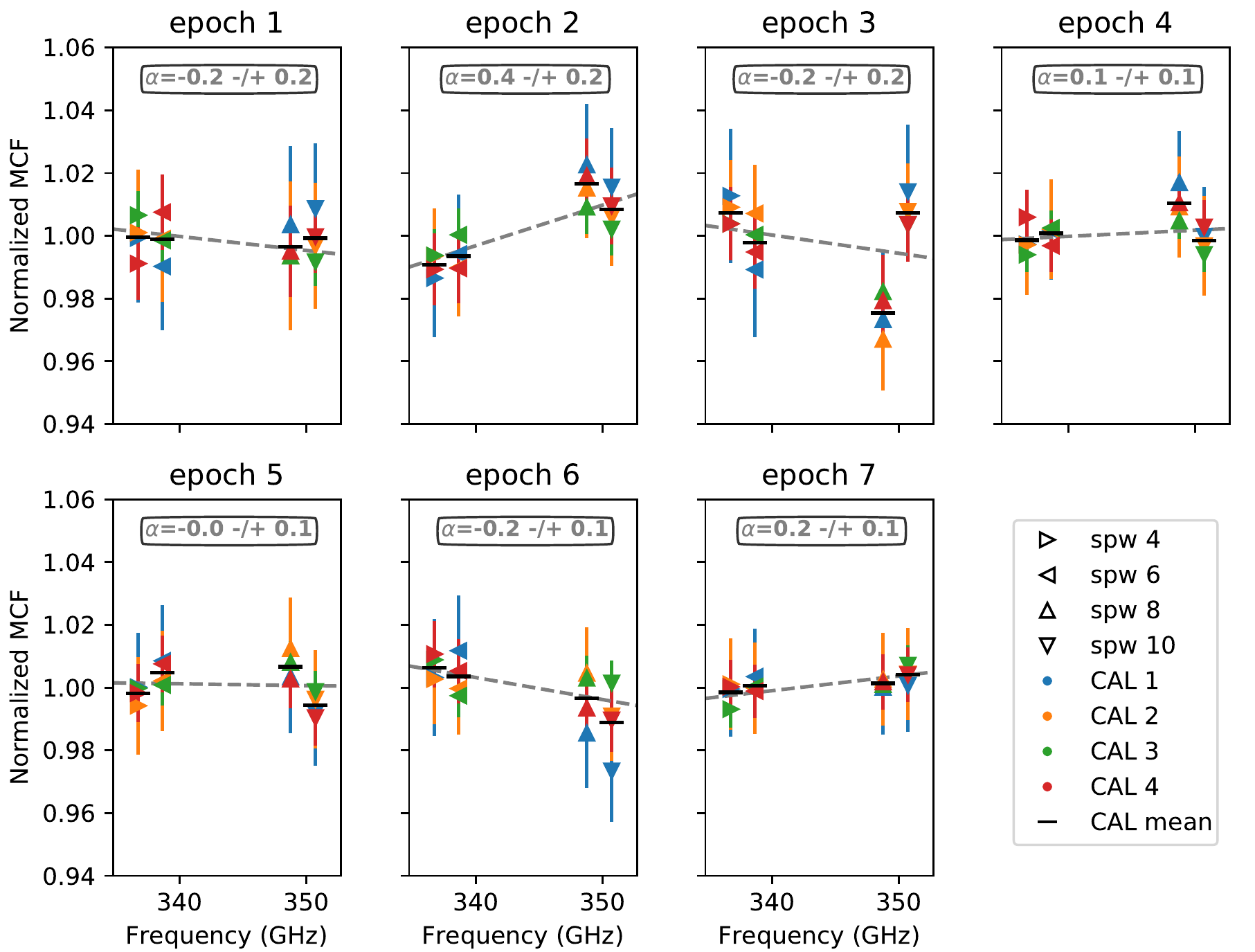}
    \caption{Mean correction factors (MCFs) per spectral window normalized to the MCF from all spectral windows (colored triangles) vs observing frequency for each epoch and YSO calibrator. The average across calibrators per spectral window and epoch is marked with a black bar. }
    \label{fig:alma_norm_MCFs}
\end{figure*}

In \deleted{the seven upper-left panels of} Figure \ref{fig:alma_norm_MCFs}, we show the MCF for each YSO calibrator and spectral window normalized to the \replaced{rFCF obtained for each epoch}{MCF obtained using data from all spectral windows}, where the error bars are computed \replaced{from the uncertainty in the point source fits}{using equation \ref{eqn:mcf_err} with normalization treated as a constant}. The magnitude of the normalized MCFs is small but correlated across calibrators, implying an additional source of uncertainty in the relative flux calibration between spectral windows. \added{We note that we obtain similar results before and after the self-calibration of our data in section \ref{ssec:self-cal}.}

Flux calibration errors between spectral windows may be systematic if there is a frequency dependence, which could potentially occur if the \added{pipeline generated} spectral index of the grid calibrator was incorrect. We thus fit a power law \added{of the form $C\nu^\alpha$} to the normalized MCFs vs frequency \added{using the function \texttt{optimize.curve\_fit} in the \texttt{scipy} python package} and assuming that the uncertainty in the normalized MCFs is entirely due to the point source fitting, and find that all but epoch 2 have a power law index consistent with zero.\added{\footnote{We have also performed the fits using a residual bootstrapping procedure incorporating monte carlo treatment of the noise, which can provide more robust error estimates for small data sets. We find the bootstrapping slopes and uncertainties are similar to those from \texttt{optimize.curve\_fit}, except for epoch 6, where the error bars are much larger ($\alpha=-0.2^{+0.2}_{-0.9}$).  }}
\added{Thus, in only one epoch there is evidence of a residual frequency dependence in the flux calibration.}
Checking for variability in the pipeline spectral index near the date of epoch 2, we find that the original and updated spectral index of $J1751+0939$ are $-0.590\pm0.037$ and $-0.482 \pm 0.014$, both of which are reasonably consistent with historical measurements (see Section \ref{sec:spix_epoch2} of the Appendix). If we correct the epoch 2 updated spectral index of $J1751+0939$ using the value of the normalized MCFs, the true spectral index would be $\sim 0$. This value is extremely inconsistent with monitoring of $J1751+0939$, and moreover, such an index is unlikely for a quasar, as the quasar brightness at mm wavelengths is dominated by synchrotron emission \replaced{\citep{vankempen2014}}{\citep{vankempen2014,planck2011}}. A large systematic error introduced by an incorrect quasar spectral index is therefore ruled out for epoch 2. 

\replaced{
We now consider if variation in the normalized MCFs is instead due to random error in the relative flux calibration of spectral windows, which could explain epochs where there is a relative offset between all spectral windows and no systematic frequency dependence (e.g epochs 3 and 4). The left panel of Figure \ref{fig:alma_norm_MCFs} shows the histogram of the normalized MCFs, which is well-described by a Gaussian fit with $\sigma=1.12\%$. The width of this distribution can be described as the sum of two random errors: those introduced from the point source fits  and those from a relative calibration error between spectral windows. In the middle panel of Figure \ref{fig:alma_norm_MCFs}, the blue histogram shows the scatter in the normalized MCFs where the mean across the 4 calibrators has been subtracted for each spectral window and epoch, while the red histogram shows the distribution of these subtracted mean values. A Gaussian fit to the blue histogram has a width of $\sigma=0.52\%$, which is approximately the same as the typical uncertainty in our point source flux measurements for a single spectral window. A Gaussian fit to the red histogram has a width of $\sigma=0.99\%$, which we identify as the magnitude of the relative flux calibration error between spectral windows. 
}
{
We now consider if variation in the normalized MCFs is instead due to random error in the relative flux calibration of spectral windows, which also could explain epochs where there is a relative offset between all spectral windows and no significant systematic frequency dependence (e.g epochs 3 and 4). The left panel of Figure \ref{fig:alma_norm_MCFs_hists} shows the histogram of the normalized MCFs, which is well-described by a Gaussian fit with $\sigma=0.9\%$. We assume the width of this distribution can be described as the sum of two uncorrelated random errors: those introduced from the point source fits in the calculation of the normalized MCFs and those from a relative calibration error between spectral windows; systematic frequency dependent contributions are assumed to be negligible. In the center panel of Figure \ref{fig:alma_norm_MCFs_hists}, the blue histogram shows the scatter in the normalized MCFs where the mean across the 4 calibrators per spectral window and epoch (the black bars in Figure \ref{fig:alma_norm_MCFs}) has been subtracted  i.e.:

\begin{figure*}[htb]
    \centering
    \includegraphics[scale=0.95]{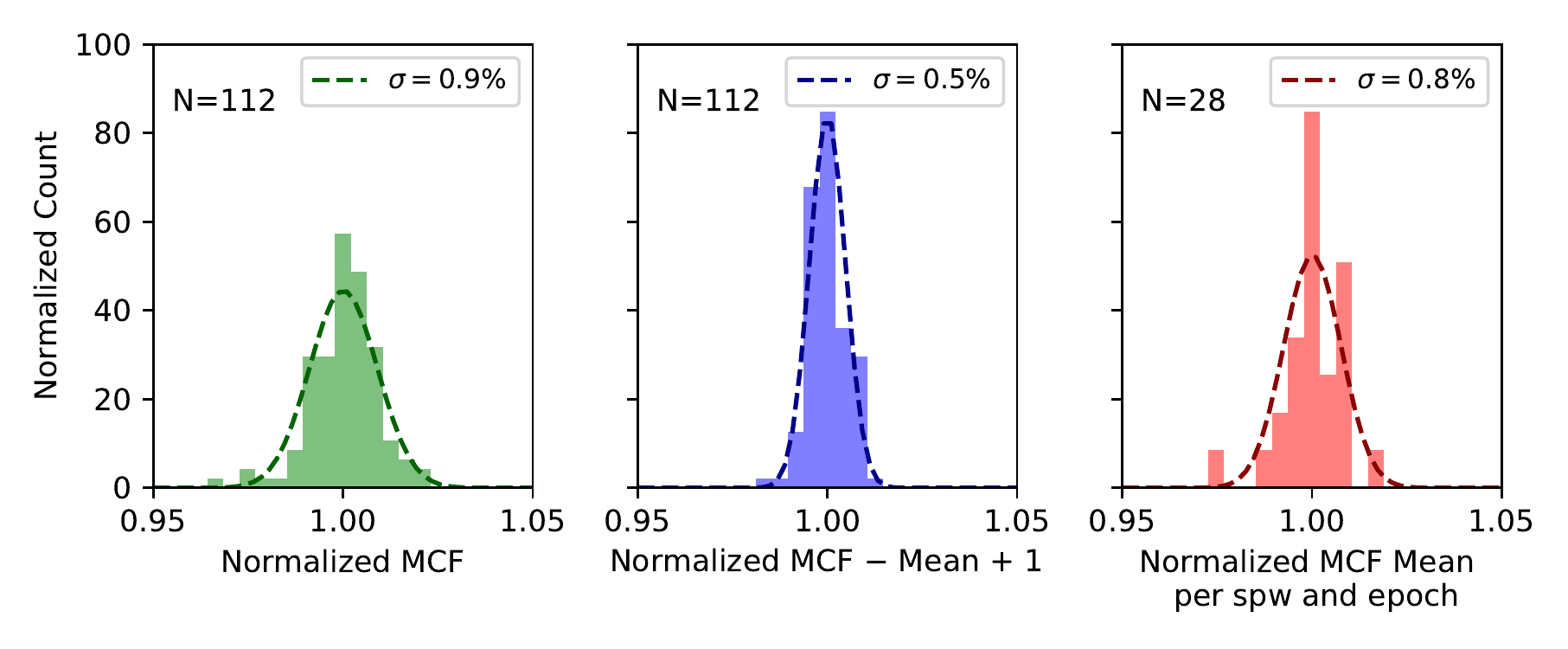}
    \caption{Left panel: histogram of the normalized MCFs in for all epochs and calibrators. Center panel: histogram of all normalized MCFs with the mean across the 4 calibrators in each spectral window subtracted. Right panel: histogram of the distribution of the subtracted mean values per epoch.}
    \label{fig:alma_norm_MCFs_hists}
\end{figure*}

\begin{equation}
\left ( \mathrm{MCF}_{i,j,k} -  \sum\limits_{j=1}^{4} \mathrm{MCF}_{i,j,k}/4  \right )/\mathrm{MCF}_{i,j}, 
\end{equation}
where $k$ is the spectral window. A Gaussian fit to the blue histogram has a width of $\sigma=0.5\%$, which is approximately the same as the typical uncertainty in our point source flux measurements for a single spectral window. In the right panel of Figure \ref{fig:alma_norm_MCFs_hists}, the red histogram shows the distribution of the subtracted mean values, i.e.:

\begin{equation}
\frac{\sum\limits_{j=1}^{4} \mathrm{MCF}_{i,j,k}/4}{\mathrm{MCF}_{i,j}}. 
\end{equation}

A Gaussian fit to the red histogram has a width of $\sigma=0.8\%$, which we identify as the magnitude of the relative flux calibration error between spectral windows. 
}

\begin{figure}[t]
    \centering
    \includegraphics[scale=1.0]{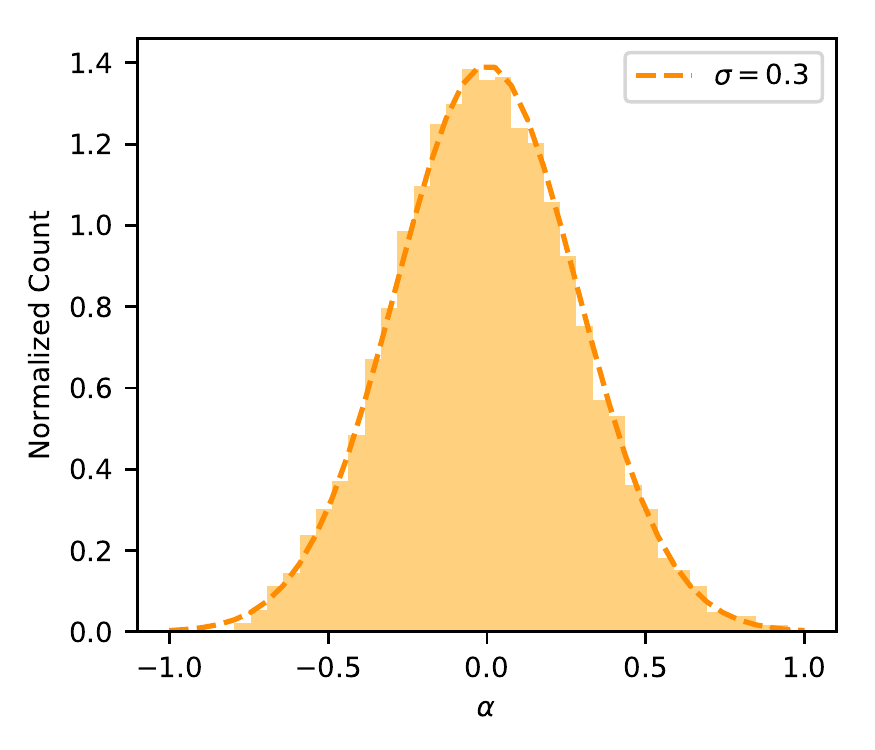}
    \caption{Monte Carlo simulation with N=10000 samples of the value of normalized MCF spectral index ($\alpha$) we should find with a 0.8\% flux scale error between spectral windows in our Band 7 observations.}
    \label{fig:mcalpha}
\end{figure}

This additional source of uncertainty between spectral windows would imply that the significance of the spectral index in epoch 2 \replaced{($4.2\sigma$)}{($2\sigma$)} is overestimated, and may simply be the result of outlier values in the relative calibration of spectral windows. We thus run a Monte Carlo simulation to generate sets of 16 normalized MCFs according to the sum the random errors from the flux measurement and relative calibration between spectral windows. We then measure the power law index $\alpha$ for each simulated set of normalized MCFs and repeat this process 10000 times. The resulting distribution of $\alpha$ has a standard deviation of \replaced{$\sigma=0.28$}{$\sigma=0.3$} and is shown in Figure \ref{fig:mcalpha}. The probability of obtaining \replaced{$\alpha=0.67$}{$\alpha>=0.4$} from random errors is \replaced{$2.4\sigma\sim 2\%$}{$\sim 2.2\%$} for one observation or $\sim16\%$ for seven, and thus the slope in the second epoch is plausibly explained as the result of a relative calibration error of \replaced{$\sim 1\%$}{$\sim 0.8\%$} between spectral windows. This relative error implies an additional source of uncertainty when comparing source fluxes between spectral windows, the impact of which is discussed further in Section \ref{ssec:best_practice}.

\section{Discussion}
\label{section:disc}

The preceding analysis has shown that: 1) without the most up-to-date calibrator catalog, the relative flux calibration accuracy of delivered ALMA data may be larger than the nominal 10\%; 2) within a single ALMA execution block in one band, there exists a \replaced{$\sim1\%$}{$\sim0.8\%$} flux calibration uncertainty between spectral windows. We now discuss the impact of these two points on various science goals and how a typical ALMA user can address them, and provide some suggestions for obtaining optimal flux calibration accuracy.

\subsection{Impact of Flux Calibration Accuracy}
\label{ssec:impacts}

The accuracy of the original pipeline flux calibration identified from the range of rFCF magnitudes is a particular concern for time domain science cases that require measurement of changes in source flux smaller than a factor of a few times the calibration accuracy. As an example, if we naively compared a single source between the outlier 2nd and 7th epochs \added{before catalog updates or self-calibration}, we would see a $\sim 45\%$ change in flux. Assuming that the band 7 ALMA calibration accuracy is  $\sim10\%$, as is stated in the ALMA documentation \citep{almaguide} and often assumed in the literature, we would \added{mistakenly} identify this as a robust detection of variability. 

A \replaced{$\sim1\%$}{$\sim0.8\%$} flux calibration uncertainty between spectral windows strongly affects the accuracy of in-band spectral index measurements due to the short length of the frequency ``lever arm''. A brief example of measuring a spectral index with various ALMA settings is illustrative. Consider observations of a source using the default ALMA spectral window frequencies for continuum observations in Bands 3 and 7, shown in Table \ref{tab:alma_def_spw}. The absolute uncertainty of a spectral index measured between frequencies $\nu_1$ and $\nu_2$ is $\sigma_\alpha = \sqrt{2}\sigma_{F}/\ln(\nu_2/\nu_1)$, where $\sigma_{F}$ is the relative flux uncertainty and $\nu_2 > \nu_1$. Assuming \replaced{$\sigma_{F}$=1\%}{$\sigma_{F}$=0.8\%}, the uncertainty in the spectral index comparing spectral windows 1 and 4 is thus \replaced{0.10}{0.08} for Band 3 and \replaced{0.35}{0.28} for Band 7. For comparison, a spectral index measured between spw 1 in band 3 and spw 4 in band 7 with the nominal $\delta_{F}$=10\% would have an uncertainty of 0.01. These are only lower limits on the expected uncertainties, as in reality any flux measurement will have additional uncertainties from the model fitting. Even a small relative flux calibration error between spectral windows is therefore problematic for measurement of in-band spectral index at the higher ALMA frequencies.

\begin{deluxetable}{cccccc}
\label{tab:alma_def_spw}
\tablecaption{ALMA default continuum spectral window frequencies.}
\tablehead{\colhead{Band} & \colhead{spw 1} & \colhead{spw 2} & LO1 & \colhead{spw 3} & \colhead{spw 3} \\ & (GHz) & (GHz) & (GHz) & (GHz) & (GHz) }
\startdata
3 & 90.5& 92.5 &  97.5& 102.5& 104.5\\
4 &138.0& 140.0& 145.0& 150.0& 152.0\\
5 &196.0& 198.0& 203.0& 208.0& 210.0\\
6 &224.0& 226.0& 233.0& 240.0& 242.0\\
7 &336.5& 338.5& 343.5& 348.5& 350.5
\enddata
\tablecomments{LO1 is the local oscillator frequency.}
\end{deluxetable}

In general, underestimating the flux calibration accuracy is a problem for science goals where this is the limiting factor in the analysis. A recent example is the modelling of millimeter-scattering processes in the TW~Hya protoplanetary disk \citep{ueda2020}. The authors fit radiative transfer models with and without scattering to SEDs of the object obtained in ALMA bands 3, 4, 6, 7, and 9. Both models fit the data within the uncertainty of the flux measurements, which were dominated by the flux calibration accuracy. While \cite{ueda2020} carefully checked the variability of their calibrators and consequently adopted larger than nominal uncertainties, typical publications containing ALMA data assume the nominal uncertainties in their interpretation. Careful analysis is recommended for any case where the significance of the results strongly depends on the calibration accuracy.

\subsection{Best Practices for ALMA Flux Calibration}
\label{ssec:best_practice}

We have found that an out-of-date calibrator catalog can increase the flux calibration uncertainty above the nominal ALMA values. For any ALMA observation, it is thus worth ensuring that the catalog used by the pipeline is up-to-date. An ALMA user can compute the flux density of a grid calibrator using the same procedure as the pipeline with the function \texttt{getALMAFlux} in the \texttt{analysisutils}\footnote{\url{https://casaguides.nrao.edu/index.php?title=Analysis_Utilities}\\ \url{https://safe.nrao.edu/wiki/bin/view/Main/CasaExtensions}} python package. If the flux computed with \texttt{getALMAFlux} differs from the pipeline value, additional measurements close to the date of observation have likely been added or updated. We find that the ALMA catalog should in general be stable after a month (see Section \ref{sec:alma_monitoring} of the Appendix), so an ALMA user requiring the most accurate absolute calibration should check for catalog changes a month after the science observation. 

Changes to the flux calibrator values should also be checked for consistency with the calibrator light curves\footnote{available at \url{https://almascience.eso.org/sc/}}. In principle, the phase calibrator can also be used for a secondary consistency check, however, this is difficult as the phase calibrators are also variable quasars, are monitored infrequently and are often fainter (see \ref{sec:phasecal_check} of the Appendix). 

Users examining the pipeline weblog to check calibrator fluxes should be cautious of \replaced{the following:}{interpreting} the derived quantities for calibrators presented in tabular form on the \texttt{hifa\_gfluxscale} page \added{as flux densities}, \deleted{should not in general be interpreted as flux densities} because this is only true in the limit of high SNR.  Although these quantities have units of Jy, they are merely scale factors from the calibration table, and will be biased upwards in cases of low SNR and/or decorrelation. Nevertheless, when these factors are applied to the visibility data in the later stage \texttt{hif\_applycal}, they will yield (except in extreme cases of low SNR) calibrated amplitudes that represent the correct flux density and will produce an image of a point source with the correct flux density\footnote{The next ALMA pipeline release (2020.1.0) will now also show the mean
calibrated visibility amplitude in the \texttt{hifa\_gfluxscale} weblog table, which is
usually a very good match to flux density in the subsequent calibrator image.}.

\replaced{If}{Once} the calibrator catalog has been updated, an ALMA user can re-scale their visibility amplitudes using the {\bf\texttt{applycal}} task in CASA. Alternatively, the values in the \texttt{flux.csv} file used in stage 1 of the pipeline can simply be modified and the pipeline re-run with that file present in the working directory.

If better than the nominal ALMA flux calibration is desired, several strategies should be considered, depending on whether the relative scaling between observations or the absolute accuracy \replaced{which  matters}{is of greater importance}. For relative scaling of observations at the same frequency, a good model of one or more bright and stable science targets can be used to re-scale the visibility amplitudes using ratios of the model flux between epochs. \added{Phase self-calibration of the science targets is important to carry out in order to reduce the effect of varying phase noise on the flux scaling between epochs.} For science goals where time-variability of the sources is of interest, additional stable objects should be added as ``science'' targets in Phase 1 of the ALMA Observing Tool as we have done for our ALMA Serpens protostar variability projects. This strategy allows us to reach a relative flux calibration accuracy of $\sim 3\%$ which if reproduced for other projects, would enable science goals not possible with ALMA's nominal flux calibration accuracy.
The quasar CHECK sources automatically added to long-baseline ($\theta_{\rm beam} < 0.25''$) and high frequency ($>385$ GHz) observations by the Observing Tool (used by the pipeline to assess astrometric accuracy and phase and amplitude transfer) are too faint to rely on for the purpose of rescaling observations, are not guaranteed to be the same object between executions, and are themselves variable. For observations of a time-variable spectral line against a constant continuum, an ``in-band calibration'' strategy requiring no extra calibrators has been successfully used for monitoring of the carbon star  IRC +10216 \citep{he2019}\added{, and a similar technique was used to show a robust change in H$^{13}$CO+ line flux of the IM Lup protoplanetary disk by \cite{cleeves2017}}. Surveys observing the same field repeatedly at a given frequency will benefit from using relative calibration to re-scale the visibilities of individual execution blocks, as this will reduce artefacts in deep images and improve the self-calibration solutions. A variant on this strategy was used by the DSHARP survey, wherein a model-free approach exploiting the inherent $uv$-plane symmetry of disk sources was adopted \citep{andrews2018}. 

% NOTE: Todd added the following paragraph
For spectral scans, if the tunings are split between schedule blocks, they might be executed with different calibrators and might be executed weeks or months apart.   For this reason, it is beneficial to include a short observation of a grid source near \replaced{your}{the} science target as an additional science target in order to be sure that you have a common source with which to test the consistency of the flux calibration across executions and apply corrections to the calibrated data when necessary.

Relative calibration may be helpful for comparisons of archival ALMA data to search for time-variability. However, careful analysis is needed for identification of stable reference targets for relative calibration, and for mitigating the effects of differences in $uv$-coverage and observing frequency, which is important for both the reference and science targets \citep[see][]{francis2019}.

For observations where high absolute accuracy is needed, requesting a solar system object observation is best if one is available, however, this is not possible for high-frequency and/or long-baseline observations with small synthesized beams where the solar system objects are resolved out. For such observations, a grid calibrator should be included, and additional observations with the ACA of a solar system object and the desired grid calibrator as science targets should be requested within a few days of the primary observation and at the same frequency. For analysis of archival data, a user can search for observations within a few days of the observing date in the same ALMA band which include one of the science targets or calibrators as well as a solar system object.

\section{Conclusions}
\label{sec:conclusions}

We have used ALMA observations of 4 stable YSO calibrators to independently assess the accuracy of the ALMA pipeline flux calibration between observations and spectral windows. Our \replaced{major}{primary} findings are as follows: 

\begin{itemize}
    \item Without an up-to-date catalog including all flux calibrator observations near the observing date, the ALMA flux calibration accuracy in Band 7 may be poorer than the nominal 10\%. This problem can be identified and corrected by an ALMA user using the \texttt{analysisUtils} python package.
    \item ALMA's relative flux calibration accuracy may be further worsened by phase decorrelation due to poor weather if self-calibration is not possible or not applied.
    \item We obtain a relative ALMA Flux calibration accuracy of $\sim 3\%$ with observations of four additional bright and stable YSO calibrators and simple $uv$-plane modelling. Calibration to this level of accuracy enables science goals which would not be possible within the nominal ALMA flux calibration uncertainties.
    \item We find our observations show a relative flux calibration uncertainty between spectral windows of \replaced{1}{0.8}\%, implying that measuring spectral indices within an ALMA band may be highly uncertain, e.g., with default Band 7 continuum spectral windows of bright targets, the spectral index uncertainty from in band measurement is $\sim0.3$.
    \item In light of typical ALMA observing practices and constraints, science goals requiring high flux accuracy should be performed in a manner that assures a robust calibration, such as the observation of additional calibrators combined with a relative calibration strategy, and observation of solar system objects for high absolute accuracy. 
\end{itemize}

\section{Acknowledgements}

\added{The authors appreciate the important contribution of the JCMT Transient Team members in helping to motivate our studies of sub-mm variability with ALMA.} We thank the anonymous referee for their helpful comments on this paper.
This paper makes use of the following ALMA data: ADS/JAO.ALMA\#2018.1.00917.S, ADS/JAO.ALMA\#2019.1.00475.S. ALMA is a partnership of ESO (representing its member states), NSF (USA) and NINS (Japan), together with NRC (Canada), MOST and ASIAA (Taiwan), and KASI (Republic of Korea), in cooperation with the Republic of Chile.  The Joint ALMA Observatory is operated by ESO, AUI/NRAO and NAOJ.  The National Radio Astronomy Observatory is a facility of the National Science Foundation operated under cooperative agreement by Associated Universities, Inc. DJ is supported by the National Research Council Canada and an NSERC Discovery Grant. GJH is supported
by general grant 11773002 awarded by the National Science
Foundation of China. DH acknowledges support from the EACOA
Fellowship from the East Asian Core Observatories Association. 

\software{astropy \citep{astropy},
          matplotlib \citep{matplotlib2007},
          Common Astronomy Software Application (CASA) 5.6.1 \citep{casa2007}
          }.
          
\bibliography{almacal_arxiv}{}
\bibliographystyle{aasjournal}

\appendix
\section{ALMA Flux and Phase Calibrator Light Curves}
\label{sec:almacal_lcurves}

\begin{figure*}[htb]
    \centering
    \includegraphics[scale=1.0]{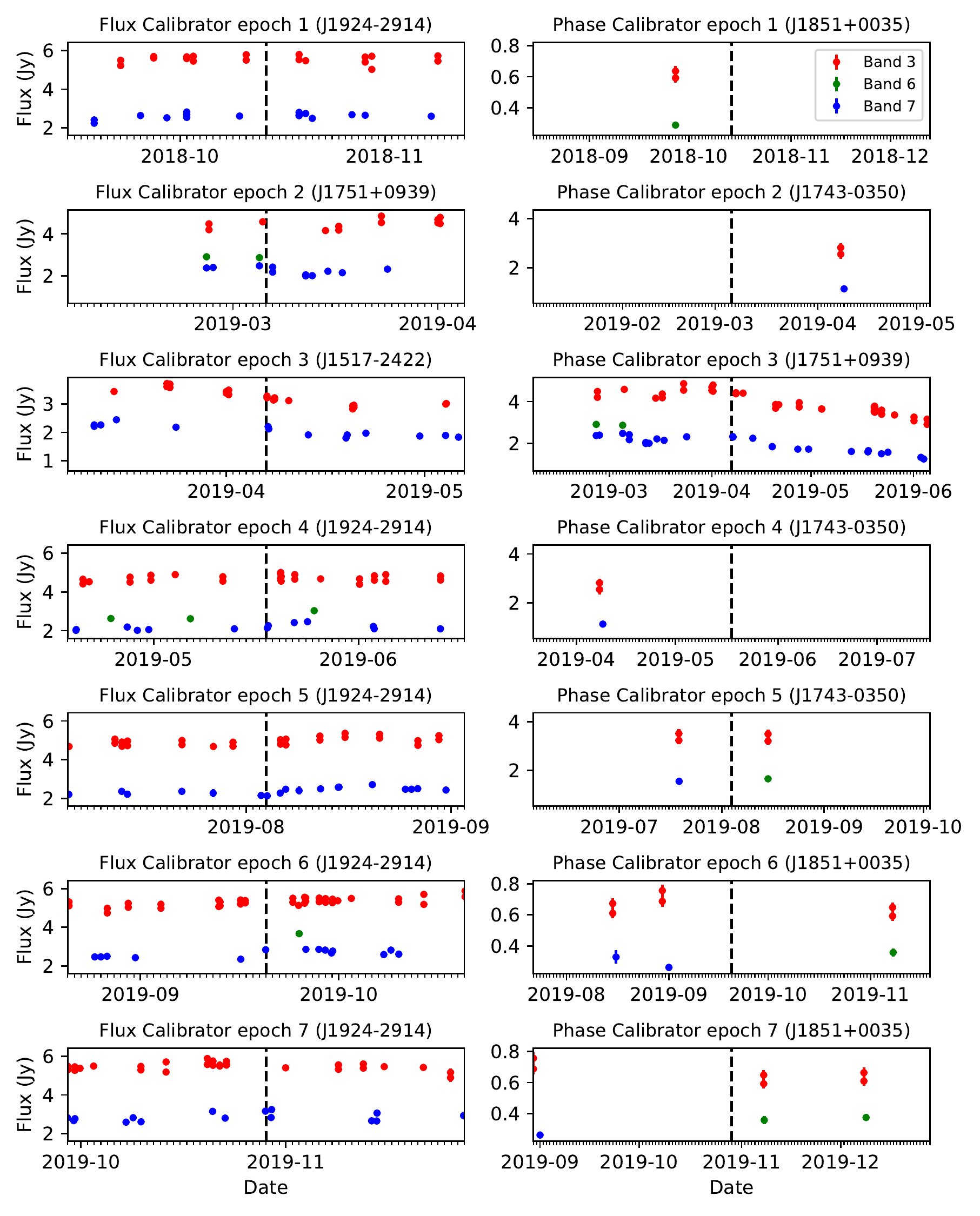}
    \caption{Catalog light curves of the grid flux calibrators (left column) and phase calibrators (right column) for our ACA observations, centered on the observing dates (dashed line). Band 3, 6, and 7 observations are shown as red, green, and blue markers respectively. Flux measurements for Band 3 are made separately for the upper and lower sideband. The light curves are shown with a 2 month range around the observing date for the Flux calibratiors and 4 months for the less frequently monitored phase calibrators.}
    \label{fig:alma_cal_lcurves}
\end{figure*}

In Figure \ref{fig:alma_cal_lcurves}, we show the catalog light curves of the grid flux calibrators and phase calibrators (see Table \ref{tab:aca_obs}) used for our observations centered around each observing epoch. The upper and lower sidebands of the band 3 observations are recorded separately in the catalog as the frequency difference between sidebands is $> 10\%$ of the typical observing frequency, and the calibrators are bright enough to have high S/N in both sidebands. 

\section{Spectral Index of epoch 2 Flux Calibrator}
\label{sec:spix_epoch2}

\begin{figure}[htb]
    \centering
    \includegraphics[scale=1.0]{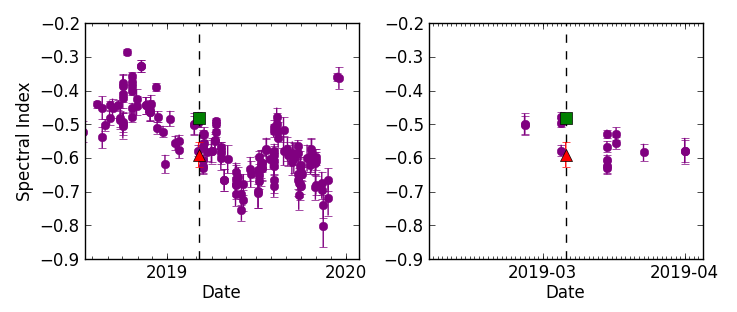}
    \caption{Catalog (purple circles), original pipeline (red triangle), and updated (green square) pipeline spectral Index of the flux calibrator J1751+0939 used in epoch 2. The date of the second epoch of ACA observations is overlaid with a black dashed line. The right panel shows a zoom-in on the second epoch within an interval of two months.}
    \label{fig:epoch2_fluxcal_spix}
\end{figure}

In Figure \ref{fig:epoch2_fluxcal_spix}, we show catalog spectral index measurements for the $J1751+0939$, the grid calibrator for epoch 2, with the date of our observations and the original and updated pipeline spectral index values overlaid. 

\section{Checking Consistency of the ALMA Flux and Phase Calibrator Flux Scales}
\label{sec:phasecal_check}

In principle, if the phase calibrator used by ALMA has been recently observed, a rFCF can be computed using the ratio of the catalog flux to the pipeline flux of the phase calibrator. As the phase calibrators are also variable quasars and are typically less frequently monitored, these rFCFs are unlikely to be any better than the grid calibrator scaling, but a large value may suggest a poor flux calibration. On the other hand, it is not generally possible to use the phase calibrator to compute the normalized MCFs used to identify differences in scaling between spectral windows (Section \ref{ssec:norm_mcfs}) as the phase calibrators typically have lower S/N than our YSO calibrators. 

\begin{figure}[htb]
    \centering
    \includegraphics[scale=0.95]{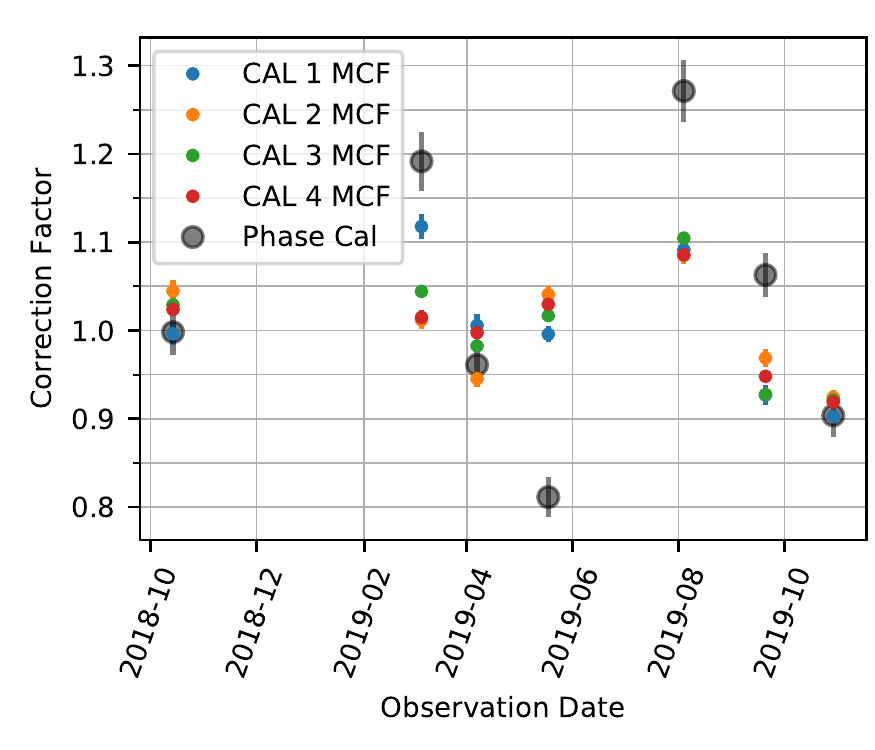}
    \caption{MCFs vs date for each of our YSO calibrators using visibilities from all spectral windows and assuming a flat spectral index. The pipeline flux calibrator values have been updated and self-calibration has been applied, as in Figure \ref{fig:alma_MCFs}. Here we also show an estimate of the rFCF using only the phase calibrator.}
\label{fig:alma_phasecal_mcfs}
\end{figure}

In  figure \ref{fig:alma_phasecal_mcfs}, we compare the MCFs computed using our YSO calibrators and the updated pipeline flux calibration with the rFCFs calculated using the phase calibrator alone. In 3 of 7 epochs, the phase calibrator rFCF agrees well with the YSO rFCF, but is inconsistent for the other 4. In comparing with the light curves in Figure \ref{fig:alma_cal_lcurves}, there is no clear relationship of a shorter delay between observation of our YSO calibrator and the phase calibrators with having a correct rFCF, except in the case of epoch 3 where a grid source observed within a week was used as a phase calibrator.

\section{ALMA Monitoring Cadence and Catalog Ingestion Delay}
\label{sec:alma_monitoring}

Using tools in \texttt{analysisUtils}, we find that the ALMA calibrator catalog entries made over the past several years typically have a delay between observation and ingestion into the catalog.  The mean value is 2-3 days, with the 90th percentile value being $\approx$1 week and a maximum value of 2 months. In Table \ref{tab:delays}, we show the delays for our the Flux calibrators used for our ACA observations of variable protostars.

\begin{deluxetable}{cccccc}
\label{tab:delays}
\tablecaption{Flux Calibrator Catalog Ingestion Delay}
\tablehead{\colhead{Epochs} & \colhead{Flux Calibrator} & \colhead{Median Lag (days) } & \colhead{90th Percentile Lag (days)} & \colhead{Maximum Lag (days)}}
\startdata
1,4,5,6,7 & J1924-2914 &  2.0 & 8.0 & 86 \\
2 & J1751+0939 &  2.0 & 9.0 & 64 \\
3 & J1517-2422 &  2.0 & 8.0 & 168\\
\enddata
%\tablecomments{}
\end{deluxetable}

\section{Measurement Set Re-scaling in CASA}

Visibility amplitude in a CASA measurement set can be rescaled using the \texttt{applycal} task. Since \texttt{applycal} applies a calibration to the DATA column and stores the calibrated visibilities in the CORRECTED column, the \texttt{split} task should first be used to create a new measurement set containing only the data to be rescaled in order to avoid overwriting the corrected column. A calibration table with the necessary complex gain factors can then be created using the \texttt{gencal} task and applied. The below python script shows an example of increasing the visibility amplitudes by 10\% which has been tested for CASA 5.6.1.  In this example, the DATA column is used because it contains the calibrated data, that is, this measurement was generated by a previous run of \texttt{split} (or \texttt{mstransform}) that pulled from the CORRECTED column.

\begin{verbatim}
# Relative change to visibility amplitude, in this case an increase of 10%.
rescale_factor = 1.1

# Split out data
split(vis=`original_data.ms,
      datacolumn=`DATA',
      outputvis=`rescaled_data.ms')

# Generate calibration table with complex gain factors to produce the desired rescaling.
gencal(vis=`rescaled_data.ms',
       caltype=`amp',
       caltable=`rescale.cal', 
       parameter=[1.0/np.sqrt(rescale_factor),])

# Apply calibration table.
applycal(vis=`rescaled_data.ms',
         gaintable=`rescale.cal')
\end{verbatim}

\listofchanges

\end{document}